\documentclass[11pt,fleqn,twoside]{article}

\usepackage{graphicx}
\usepackage{times}
\usepackage{natbib}
\usepackage{amsmath}
\usepackage{amssymb}
\usepackage{amsthm}
\usepackage{mathrsfs}
\usepackage[english]{babel}
\usepackage[all]{xy}

\renewcommand{\d}{\mathrm{d}}

\def\aap{A\&A}
\def\apj{ApJ}

\def\apjl{ApJ}
\def\mnras{MNRAS}

\def\physrep{Phys. Rep.}
\def\nat{Nature}

\def\prd{Phys. Rev. D}

\def\fnl{$f_{\rm NL}$}
\def\Mpch{{\rm Mpc/{\it h}}}

\begin{document}

\title{\Large Non-Gaussianities, early Universe, and GRBs.\\
Tracing the primeval state of the Universe with number counts of Gamma-Ray Bursts
}
\author{Umberto Maio\\
{\small Max Planck Institute for Extraterrestrial Physics}
}
\date{}
\maketitle

\section*{Abstract}
We investigate the effects of primordial non-Gaussianities in the primordial Universe and the effects on the baryonic structure formation process. By relating the cosmic star formation rate in Gaussian and non-Gaussian scenarios to the detectability of high-redshift sources of reionization, we derive the expected Gamma-Ray Burst rate in the different models.
We find that counts of high-redshift Gamma-Ray Bursts can be used as cosmological probes of non-Gaussianities and that they are suitable candidates to distinguish non-Gaussian effects at early epochs.

\section{Introduction}
Primordial density fluctuations grow during cosmological times and lead to the observed structures in the Universe \cite[e.g.][]{BarkanaLoeb2001}, via gas collapse and condensation in the evolving dark-matter haloes.
The statistical distribution of the primordial fluctuations is still debated, though.
In fact, despite the common assumption of Gaussianity, recent analyses show evidences for positively skewed distributions -- see the summary Table~2 in \cite{MaioIannuzzi2011}.
\\
High redshift represents an interesting regime to potentially investigate these non-Gaussian effects, since very early structures and primordial mini-haloes (hosting the first bursts of star formation, spreading the first heavy elements in the Universe, seeding the first black holes, being responsible for primordial cosmic re-ionization and hydrogen 21-cm emission) are expected to be somehow affected by the underlying matter distribution \cite[][]{Maio2011cqg, MaioKhochfar2012,PetkovaMaio2012}.
\\
In this respect \cite[][]{Maio2011cqg,MaioGRBsNonG2012}, a key tool for studies of high-redshift environments
might be represented by $\gamma$-ray bursts (GRBs), powerful explosions emitting $\gamma$ rays.
These bursts have (comprehensive reviews are given in \cite{Piran2004, Meszaros2006}):
isotropic equivalent peak luminosities as high as $\sim 10^{54}$ erg s$^{-1}$; an isotropic angular distribution; and a bimodal duration distribution, made mostly of long GRBs (with a period longer than 2~seconds), and in minor part of short GRBs (detected at low redshift with a period shorter than 2~seconds).
In the next, we will only consider long GRBs (LGRBs), because they are linked to the death of massive stars and hence are indicators of the local star formation episodes \cite[e.g.][]{Nuza2007, Campisi2011, Mannucci2011}.
\\
Here, we will show that GRBs can be used as cosmological probe of the amount of non-Gaussianity present in the primordial density field, because they are sensitive to the underlying cosmological model through the cosmic star formation history -- further datails on this subject can be found in \cite{MaioGRBsNonG2012}.
\\
We will reach this conclusion by performing a detailed analysis of the GRB rate in different Gaussian and non-Gaussian scenarios, based on N-body, hydrodynamic, chemistry simulations of early structure formation \cite[][]{MaioIannuzzi2011}.
The star formation rate (SFR) and the adopted initial mass function (IMF) for the stellar populations tracked during the runs are used to determine the expected GRB formation rate density in the various cases, and hence the integrated GRB rate ($R$), for both metal-poor population~III (hereafter, popIII) regime and metal-enriched population~II-I (hereafter, popII-I) regime.

\section{Model and results}\label{sect:results}
Given the cosmic SFR, $\dot\rho_\star$, tracked by the different runs as a function of $z$ in different cosmological models \cite[]{Maio2010, MaioIannuzzi2011}, the expected redshift distribution of observed GRBs can be computed once the GRB luminosity function (LF) and the GRB formation history have been specified \cite[e.g.][]{Salvaterra2007, Salvaterra2009, Salvaterra2012CompleteSwiftSample}.
From these, we then compute the evolution of the GRB formation rate
density, $\dot n_{\rm GRB}$, and the corresponding integrated GRB rate, $R$.
\\
The (comoving) SFR densities are obtained by a set of numerical N-body, hydrodynamical, chemistry simulations with two different box sizes (0.5 and 100~Mpc/{\it h}, with $h=0.7$) starting from initial conditions with a different level of primordial non-Gaussianity, parameterized in terms of the non-linear parameter \fnl $ = 0, 10, 50, 100, 1000$.
\\
The observed peak photon flux, $P$, emitted by an isotropically radiating source at redshift $z$ and corresponding luminosity distance $d_L(z)$, as detected in the energy band $E_{\rm min} < E < E_{\rm max}$, is
\begin{equation}
P=\frac{(1+z)}{4\pi d_L^2(z)}~ \int^{(1+z)E_{\rm max}}_{(1+z)E_{\rm min}}\!\!\! S(E)~\d~E,
\end{equation}
\noindent
where $S(E)$ is the differential rest-frame photon luminosity of the
source.
To describe the typical burst spectrum we adopt a ``Band'' 
function with low- and high-energy spectral indices equal to $-1$ and
$-2.25$, respectively, and normalization:
\begin{equation}
L=\int^{10\,\rm{MeV}}_{1\,\rm{keV}} E ~S(E) ~\d E,
\end{equation}
with $L$ isotropic-equivalent peak luminosity.
\\
Given a normalized GRB LF, $\psi(L)$, the observed number rate of bursts (in $\rm yr^{-1}$) at redshift $z$, with peak photon flux, $P$, between $P_1$ and $P_2$ is
\begin{eqnarray}
\label{eq:Ndot}
\dot N(z) 
\equiv \frac{\d N_{P_1<P<P_2}(z)}{\d t}  
= \int_z^{\infty} \d z^\prime ~\frac{\d V(z^\prime)}{\d z^\prime} ~\frac{\dot n_{\rm GRB}(z^\prime)}{(1+z^\prime)} \times \int^{L_{P_2}(z^\prime)}_{L_{P_1}(z^\prime)} \psi(L^\prime) \d L^\prime,
\end{eqnarray}
where the factor $(1+z^\prime)^{-1}$ accounts for cosmological time dilation, $V(z')$ is the comoving volume at redshift $z'$, and $\dot n_{\rm GRB}$ is the actual comoving GRB formation rate density.
\\
By assuming that  GRBs are good tracers of star formation, we can write that the GRB formation rate density is proportional to the SFR density:
\begin{equation}
\dot n_{\rm GRB}(z) \equiv k \dot{\rho}_\star(z),
\end{equation}
where the normalization constant, $k$ (whose dimensions are the inverse of a mass), incorporates further not-well-known effects,
like GRB beaming, efficiencies, and black-hole production probability (depending on the adopted IMF).
\\
As for the LF, we will adopt a normalized GRB LF described by a single power-law with slope $\xi$ and
decreasing exponentially below a characteristic luminosity, $L_{\star}$,
\begin{equation}\label{eq:LF}
 \psi(L) \propto \left(\frac{L}{L_{\star}}\right)^{-\xi} \exp \left(-\frac{L_{\star}}{L}\right).
\end{equation}
\noindent
Additionally, we take into account possible rdshift evolution of the GRB LF by setting
$L_{\star}(z)=L_{{\star},0}(1+z)^\delta$, 
where $L_{{\star},0}$ is the characteristic luminosity at $z=0$,
and $\delta$ is the evolution parameter.
\\
For simplicity, the normalization of $\psi(L)$ is included in $k$, and it is fixed when the GRB number rate in equation~(\ref{eq:Ndot}) is normalized to the rate observed at $z=0$.
\\
From the previous relations we can finally compute the GRB rate
(in units of yr$^{-1}$~sr$^{-1}$), $R$, as the derivative with respect to the solid angle, $\Omega$, of the GRB number rate, $\dot N$:
\begin{equation}
\label{eq:R}
R(z) = \frac{\d \dot N(z)}{\d \Omega}.
\end{equation}
We determine the parameters of the model by fitting the recent {\it Swift} redshift-complete sample by \cite{Salvaterra2012CompleteSwiftSample}.
\\


\noindent
We apply our calculations to the different non-Gaussian cosmologies previously mentioned to derive count predictions as a function of $z$.
We underline that the following results are valid for an ideal instrument, that is able to detect all the GRBs at any redshift.
The main effects of primordial non-Gaussianities on GRBs are due to the differences in the redshift evolution of the SFRs in the various scenarios \cite[see][]{MaioIannuzzi2011} and, hence, in the consequently different GRB rates.
\begin{figure*}
\centering
\includegraphics[width=0.49\textwidth]{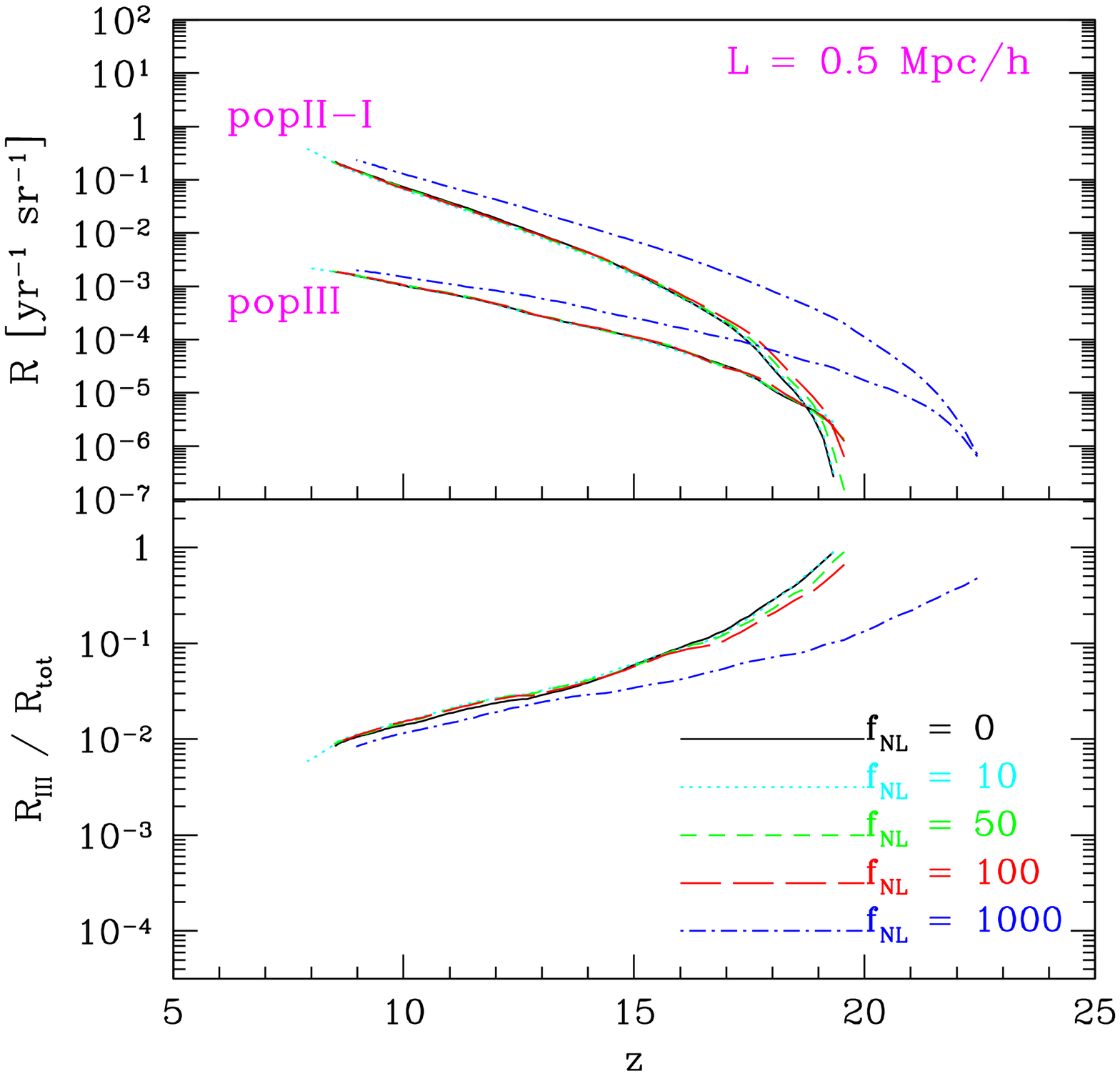}
\includegraphics[width=0.49\textwidth]{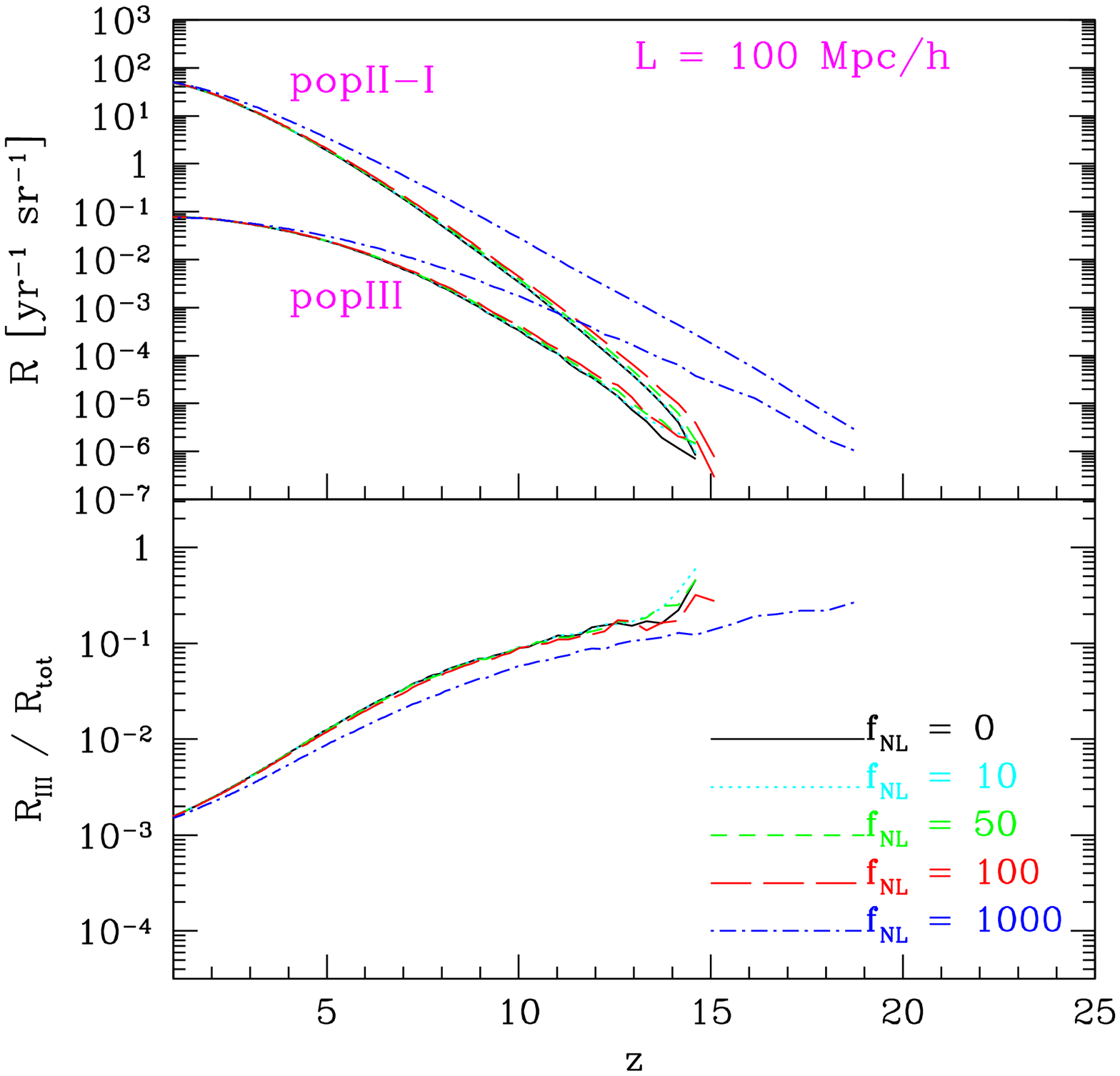}
\caption[Redshift comparison]{\small
Top panels: The expected popII-I and popIII GRB rates, $R$, in the $0.5\,\Mpch$-side boxes (left), and in the $100\,\Mpch$-side boxes (right), for models with different primordial non-Gaussianity: \fnl = 0 (solid black line), 10 (dotted cyan line), 50 (short-dashed green line), 100 (long-dashed red line), and 1000 (dot-dashed blue line).
Bottom panels: The corresponding relative contributions of the popIII GRB rates, $R_{\rm III}$, to the total rate, $R$, for the different cosmological models. See \cite{MaioGRBsNonG2012}.
}
\label{fig:Rall}
\end{figure*}

\noindent
In Fig.~\ref{fig:Rall} (see also \cite{MaioGRBsNonG2012}), we plot the GRB rate, $R$, for the small 0.5~\Mpch-size boxes (left panels) and the large 100~\Mpch-size boxes (right panels).
In the top panels, we show the redshift evolution for all the \fnl{} values considered, while in the bottom panels we focus on the relative contribution of the popIII GRB rate ($R_{\rm III}$) to the total rate, that is widely dominated by popII-I stellar generations, at redshift lower than $z\sim 20$.
\\
In both small and large volumes the effects due to the presence of primordial non-Gaussianities are visible at $z > 10-15$.
The rates eventually converge at later times, when feedback processes start dominating the gas behaviour and the resulting star formation.
\\
These trends are valid for both popII-I and popIII regimes, even
though the latter is usually negligible, predicting popIII GRB
rates, $R_{\rm III}$, that, following the behaviour of the popIII SFRs, drop by two orders of magnitude (bottom-left panel).
Further discussions can be found in \cite{MaioGRBsNonG2012}.

\section{Conclusion} \label{sect:conclusions}
Assuming that long $\gamma$-ray bursts are fair tracers of star formation \cite[][]{Nuza2007, Campisi2011, Mannucci2011}, we propose to use them as probes of the underlying matter distribution at high redshift, when the possible presence of non-Gaussianity would have visible effects on the baryon evolution (we refer the interested reader to \cite[][]{MaioGRBsNonG2012} for further details and discussions).
\\
We find that already at $z > 6$ cosmologies with large \fnl{} values have distinctive characteristics compared to those of the Gaussian case.
At very early times ($z\sim 15-20$) the boost in the rate due to non-Gaussianities is $\sim 2 - 3$ orders of magnitudes for \fnl=1000, and up to a factor of $\sim 10$ for \fnl=100, while milder differences of a factor of $\sim 2$ are still visible for values of \fnl $\sim$ 50.
These effects are particularly evident on the total GRB rate, that is largely dominated by popII-I stars, while the result for the popIII GRB rate is noisier, mostly for \fnl$\sim 0-100$, as a consequence of the corresponding, short-lived, popIII star forming regime \cite[][]{Maio2010, Maio2011b}.
We highlight that differences in the GRB rate induced by non-Gaussianities are expected to be important mostly at very high redshift, while additional hydrodynamic processes and feedback mechanisms would wash out any memory of primordial non-Gaussianities at lower $z$.
\\
The existence of GRBs at such high redshift is not unlikely, as a small population of extremely dark GRBs, i.e. bursts for which the afterglow remains undetected in spite of early and deep K~band
observations, has been recently identified \cite[][]{DEliaStratta2011}.
While the nature of these GRBs is still matter of debate, it is possible that some of them are at $z\ge 18$.
If confirmed, this could provide evidence for mildly positive non-Gaussian parameter.
To draw more definitive conclusions and give more stringent constraints much larger high-$z$ GRB complete samples, currently not available in the literature, are required.

\section*{Acknowledgements}
This work has been performed under the HPC-EUROPA2 project (project number: 228398) with the support of the European Commission - Capacities Area - Research Infrastructures.
We also acknowledge the Italian Computing center (CINECA) for kind hospitality.

\bibliographystyle{plain}

\end{document}